\documentclass{ifacconf}

\usepackage{nameref} 

\usepackage{graphicx}      
\usepackage{natbib}        
\usepackage[disable]{todonotes}
\usepackage{amsmath}
\usepackage{listings}
\usepackage{url}
\usepackage{amssymb}

\newcommand{\norm}[1]{\left\lVert#1\right\rVert}
\newcommand{\diff}[1]{\ensuremath{\operatorname{d}\!{#1}}}
\newcommand{\abs}[1]{\left\lvert#1\right\rvert}

\newcommand{\R}{\mathbb{R}}
\newcommand{\D}{\ensuremath{\mathcal{D}}} 

\begin{document}
\begin{frontmatter}

\title{Smoothness and continuity of cost functionals for ECG mismatch computation\thanksref{footnoteinfo}} 

\thanks[footnoteinfo]{This work was financially supported by the Theo Rossi di Montelera Foundation, the Metis Foundation Sergio Mantegazza, the Fidinam Foundation, the Horten Foundation and the CSCS-Swiss National Supercomputing Centre production grant s1074 to SP.}

\author[KFU,BIOTECH,MUG]{Thomas Grandits}
\author[CCMC]{Simone Pezzuto}
\author[MUG,BIOTECH]{Gernot Plank}

\address[KFU]{Institute of Mathematics and Scientific Computing, University of Graz, Graz, Austria (e-mail: thomas.grandits@uni-graz.at)}
\address[MUG]{Gottfried Schatz Research Center - Division of Biophysics, Medical University of Graz, Graz, Austria (e-mail: gernot.plank@medunigraz.at)}
\address[CCMC]{Center for Computational Medicine in Cardiology,
Euler Institute, Università della Svizzera italiana,
Lugano, Switzerland, (e-email: simone.pezzuto@usi.ch)}
\address[BIOTECH]{BioTechMed-Graz, Graz, Austria}

\begin{abstract}                
The field of cardiac electrophysiology tries to abstract, describe and finally model the electrical characteristics of a heartbeat.
With recent advances in cardiac electrophysiology, models have become more powerful and descriptive as ever.
However, to advance to the field of inverse electrophysiological modeling, i.e.~creating models from electrical measurements such as the ECG, the less investigated field of smoothness of the simulated ECGs w.r.t.~model parameters need to be further explored.
The present paper discusses smoothness in terms of the whole pipeline which describes how from physiological parameters, we arrive at the simulated ECG.
Employing such a pipeline, we create a test-bench of a simplified idealized left ventricle model and demonstrate the most important factors for efficient inverse modeling through smooth cost functionals.
Such knowledge will be important for designing and creating inverse models in future optimization and machine learning methods. 
\end{abstract}


\end{frontmatter}

\section{Introduction}
\label{sec:intro}
Modern health care is increasingly pushing towards personalized approaches for an improved therapeutic outcome over standard interventions, see~\cite{corral2020digital}. 
Precision cardiology aims at individualizing computational models of the heart from patient-specific data---possibly with non-invasive data, such as cardiac imaging and the standard 12-lead electrocardiogram (ECG)---thus to provide to cardiologists an advanced tool to improve diagnosis and to optimize the therapeutic approach.

The keystone of model personalization is the identification of the parameters, which mathematically translates to an optimization problem. 
Specifically, we focus in this contribution on the case where the loss functional is solely based on the ECG.
The electrophysiological inverse problem w.r.t.~the ECG has been already considered in previous works with either derivative-free optimization (sometimes combinational), see e.g.~\cite{gillette_framework_2021,camps_inference_2021,Pezzuto2021ECG}, or gradient-based approaches, see e.g.~\cite{GranditsGEASI2021}. 
In either case, the parameter space is typically low-dimensional, with less than 20 parameters.
To further advance the field of inverse cardiac electrophysiology, it will be indispensable to also explore higher dimensional spaces and thus allow for more varied models.
However, increasing the dimensionality of the explored parameter space comes with the obvious drawback of aggravated optimization since the exploration of the parameter space will become exponentially more difficult, sometimes emphasized as the \emph{curse of dimensionality}, see~\cite{bellman2015adaptive}.
A better optimization approach relies on the computation of the local gradient of a cost function and perform best for smooth and convex functions.
It is however currently still an open question, if complex electrophysiological simulations and matching them to ECGs is a challenging problem and how smooth the resulting loss function is.

While single studies have shown the effect of different parameters on the cost functionals of such inverse problems, see~\cite{grandits_inverse_2020,yang_application_2017}, they focused on a small subset of the parameters. 
Additionally, to the knowledge of the authors, no extensive study of the effect of cardiac electrophysiological models and ECG matching techniques on the continuity and smoothness of the resulting loss function have been conducted thus far.
The problem of matching a recorded, patient-specific ECG to a model-based ECG is particularly subtle, because ECGs are typically not very smooth and affected by noise. 
Moreover, ECG amplitude and morphology may be subject to epistemic uncertainty, due poor electrode contact or other physiological aspects (respiration, epicardial fat, torso inhomogeneities) generally excluded by the modeling framework. 
Finally, ECG alignment and detection of the onset of activation is not always clear. 
All these aspects affect the \emph{metric} to evaluate the mismatch.

In this paper, we will use several time series distance functions, applied between the optimal and current generated ECGs and visually analyze the difficulty of the resulting optimization problem in terms of visual smoothness.
Evaluating cost functionals of parameter spaces with more than 3 to 4 parameters is no simple feat.
Traditionally, a dimensionality reduction algorithm is first applied, such as principal component analysis, or more recently, machine learning inspired techniques such as auto-encoders or T-SNE~\cite{hinton_stochastic_2002}. 
In this paper we will rather explore and visualize the loss function of an 8-dimensional space of important and intertwined parameters using techniques promoted for the visualization of smoothness in residual networks (ResNets)~\cite{li_visualizing_2018}.

From a modeling perspective, we built a state-of-the-art pipeline for the preparation of simulations, from data to ECG.
The reconstruction and generation of anatomical models or features, purely from imaging data (such as MRI) is a non-trivial task that many previous studies investigated, see \cite{chen_deep_2020,arevalo_arrhythmia_2016}. Here, we will study the effect of mesh resolution of the heart on the optimization problem, ranging from $2.5\,\mbox{mm}$ to $0.5\,\mbox{mm}$ mean edge length. Moreover, given the importance of anisotropic conduction in the heart, being faster in fiber longitudinal direction, we will study its effect on the ECG w.r.t.\ the fiber angles.

Concerning the electrophysiology part, there is a vast literature on various models (bidomain, monodomain, eikonal) and their physiological accuracy, see~\cite{keener_mathematical_1998,sundnes_computing_2007,jalife_basic_2011}.
The considered gold-standard model, the bidomain system, is however still very expensive to solve~\cite{vigmond_solvers_2008}.
Our study requires us to repeatedly evaluate the electrical activation several thousand times, which necessitates evaluations in the timeframe of no more than a few seconds.
We will therefore focus on the anisotropic eikonal model, which have been shown to offer an faithful approximation to the activation of the heart, while being fast to solve, see~\cite{franzone_spreading_1993}. When using such an eikonal model and prescribing a fixed membrane potential at the computed activation time, we can efficiently compute the ECGs measured at the body surface with only minor estimation errors, see~\cite{pezzuto_evaluation_2017}.


The paper is organized as follows: In Sec.~\ref{sec:methods} we present the ECG model considered in this study, and the various metrics to compare ECGs. Several numerical experiments are summarized in Sec.~\ref{sec:results} and later discussed in Sec.~\ref{sec:disc}.

\section{Methods}
\label{sec:methods}

\subsection{The ECG model}
We present here the full pipeline to generate the ECG model from anatomical data and parameters.
\subsubsection{Geometric pipeline}

\begin{figure}[tb]
    \centering
    \includegraphics[width=\linewidth]{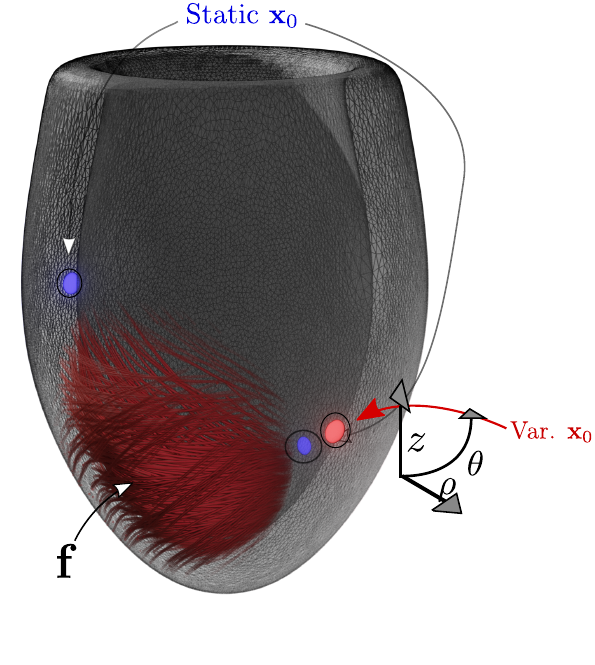}
    \caption{LV model setup with the fiber field~$\mathbf{f}$ as streamlines.
    The excitation is initiated at three different locations $\mathbf{x}_0$ at different timings: two are fixed (blue) and the last one (red) is varying.
    Spatial exploration is performed in the UVC space ($\theta, \rho, z$) that only allows feasible points inside the domain.}
    \label{fig:lv_model_w_uvc}
\end{figure}

The reference anatomy was an idealized left ventricle (LV). 
The LV was built as a volume bounded by two truncated prolate spheroidal surfaces (see Fig.~\ref{fig:lv_model_w_uvc}). 
The inner surface (endocardium) and outer surface (epicardium) have the same center and major axis, set to $37.5$~mm, whereas the minor axis are set to $41.5$~mm and $45$~mm, respectively. 
The spheroids were truncated by a plane orthogonal to the major axis and at a distance of $35$~mm from the center.
The corresponding tetrahedral mesh was obtained at different resolutions using \texttt{fTetWild}, see~\cite{hu_fast_2020}.

To ease the navigation in the LV, we make use of universal ventricular coordinates (UVCs)~\cite{bayer_universal_2018}, as depicted in Fig.~\ref{fig:lv_model_w_uvc}. The UVCs are the three functions
\[
\theta(x)\in[0,2\pi),\quad
z(x)\in[0,1],\quad
\rho(x)\in[0,1]
\]
that measure, respectively, the angular position, the normalized longitudinal distance  (from apex to base), the angular position and the normalized transmural distance (from endocardium to epicardium). Importantly, the map $\mathcal{U}: \Omega \ni x \mapsto (\theta,z,\rho)$ is invertible almost everywhere, so to make navigation in the LV easier.
This system is built by solving a sequence of Laplace equations with given Dirichlet boundary conditions, followed by a normalization (see~\cite{bayer_universal_2018}).

The fiber and sheet structure in the LV was assigned with a rule-based approach, following~\cite{bayer_novel_2012}.
Thanks to the UVCs, the local basis can be expressed as
$$
\mathbf{Q} = \begin{pmatrix} | & | & | \\ \hat{\mathbf{e}}_\theta & \hat{\mathbf{e}}_z & \hat{\mathbf{e}}_\rho \\ | & | & |  \end{pmatrix},
$$
for $\hat{\mathbf{e}}_\theta$, $\hat{\mathbf{e}}_z$, $\hat{\mathbf{e}}_\rho$ being the orthonormalized basis formed by applying Gram-Schmidt to the set $\{ \nabla\theta, \nabla z, \nabla \rho \}$. In order to define the fiber $\mathbf{f}$, sheet $\mathbf{s}$ and normal $\mathbf{n}$ direction, the local basis in $\mathbf{Q}$ was eventually rotated, first with a rotation of angle $\alpha(\rho)$ around the direction $\hat{\mathbf{e}}_\theta$, and then with a rotation of angle $\beta(\rho)$ around $\hat{\mathbf{e}}_\rho$:
\begin{equation}
\begin{pmatrix}
    \vert & \vert & \vert \\
    \mathbf{f} & \mathbf{s} & \mathbf{n} \\
    \vert & \vert & \vert 
\end{pmatrix} = \mathbf{Q}
\begin{pmatrix}
    \cos \alpha & -\sin \alpha & 0 \\
    \sin \alpha & \cos \alpha & 0 \\
    0 & 0 & 1
\end{pmatrix}
\begin{pmatrix}
    1 & 0 & 0\\
    0 & \cos \beta & \sin \beta \\
    0 & -\sin \beta & \cos \beta
\end{pmatrix}.
\label{eq:fiber_sheet_normal}
\end{equation}
The angles $\alpha(\rho)$ and $\beta(\rho)$ were linearly interpolated, as follows:
\begin{align*}
\alpha(\rho) &= \alpha_1 + (\alpha_2-\alpha_1)\rho, \\
\beta(\rho) &= \beta_1 + (\beta_2-\beta_1)\rho.
\end{align*}

\subsubsection{Electrophysiological pipeline}

We model the electrical activation throughout the heart, denoted by~$\phi(x)$, $x\in\Omega$ with the anisotropic eikonal equation
\begin{equation}
\begin{cases}
\sqrt{\mathbf{D}\nabla\phi\cdot\nabla\phi} = 1, & x\in\Omega, \\
\phi(x_k) = \phi_k, & k = 1,\ldots,N.
\end{cases}
\label{eq:eikonal}
\end{equation}
The point-wise initial conditions were set through the UVC system, that is $x_k$ is the projection to the closest mesh vertex of the point $\mathcal{U}^{-1}(\theta_k,z_k,\rho_k)$. The symmetric
positive-definite conduction velocity tensor $\mathbf{D}$, was defined as follows:
\begin{equation}
\mathbf{D} = K\cdot \mathbf{G}_\mathrm{i}(\mathbf{G}_\mathrm{i}+\mathbf{G}_\mathrm{e})^{-1}\mathbf{G}_\mathrm{e},
\end{equation}
that is, we assumed a monodomain approximation of the electric conductivity. For the sake of simplicity, we assumed $K=1$ in our experiments. The tensors $\mathbf{G}_\mathrm{i}$ and $\mathbf{G}_\mathrm{e}$, respectively denoting the intra- and extra-cellular electric conductivity, were set as follows:
\begin{equation}
    \mathbf{G}_\mathrm{i,e} = 
    \begin{pmatrix}
        \vert & \vert & \vert \\
        \mathbf{f} & \mathbf{s} & \mathbf{n} \\
        \vert & \vert & \vert 
    \end{pmatrix}
    \begin{pmatrix}
        g^{i,e}_f & & \\
        &    g^{i,e}_s & \\
        & & g^{i,e}_n
    \end{pmatrix}
    \begin{pmatrix}
        \text{---} & \mathbf{f} & \text{---}  \\
        \text{---}  & \mathbf{s} & \text{---}  \\
        \text{---} & \mathbf{n} & \text{---}  
    \end{pmatrix},
    \label{eq:intra_cond_tensor}
\end{equation}
with $g^e_f = 0.12\:\text{S}/\text{m}$, $g^e_s = g^e_n = g^i_n = 0.08\: \text{S}/\text{m}$. The values of $g^i_f$ and $g^i_s$ are variable.

For solving the eikonal equation~\eqref{eq:eikonal}, we used the \texttt{fim-python} library, see~\cite{grandits_fast_2021}, which implements the fast iterative method solver from~\cite{fu_fast_2013}.

The depolarization from of the membrane potential was approximated by shifting a characteristic waveform $v_m$ as follows:
\begin{equation}
    v_m(x, t) = k_0 + \frac{k_1-k_0}{2}\left[\tanh\left(2\frac{t - \phi(x)}{\tau_1}\right)+1\right],
    \label{eq:waveform}
\end{equation}
for $k_0 = -85\:\text{mV}$, $k_1= 30\:\text{mV}$ and $\tau_1 = 1\:\text{ms}$. Finally,
the electrical measurement for an electrode combination (referred to as leads) can then be computed as (see~\cite{keener_mathematical_1998}):
\begin{equation}
    V_j(t) = \int_\Omega \left< \nabla Z_j(x), \mathbf{G}_i(x) \nabla v_m(x, t)\right> \diff{x},
    \label{eq:ecg}
\end{equation}
where $j=1,2,3$ and $Z_j$ refers to the lead field. Here, for the sake of simplicity,
we considered $\nabla z_j$ equal to the canonical basis vectors $\mathbf{e}_j$, yielding
a pseudo axis-aligned Frank ECG~\cite[p.~529]{keener_mathematical_1998}.

\subsection{Loss functionals}
\label{sec:loss_functionals}

We consider the following problem: given a target ECG, denoted by~$\hat{V}_j(t)$ for $j=1,2,3$, we aim at identifying the free parameters in the ECG model depicted above, such that the mismatch between the target ECG and the simulated one is minimal. We discretize the time with a fixed time step $\Delta t = 1\,\text{ms}$, that is $t_n=n\Delta t$ for $n=0,\ldots,M$ and $M\Delta t = T$. The simulated and target ECGs are sampled at $t=t_n$, and the resulting vector for each lead, denoted by $\mathbf{V}_j$ and $\hat{\mathbf{V}}_j$, have respectively components $[V_j]_n = V_j(t_n)$ and $[\hat{V}_j]_{n} = \hat{V}_j(t_n)$.

\subsubsection{Parameter space}
\label{sec:parameter_space}

The free parameters in the model and their range of definition are:
\begin{itemize}
    \item $\alpha_1 \in [-130, 10]$ and $\alpha_2 \in [-10, 130]$ (in degrees).
    \item $g^i_f \in [0.08, 0.6]$ and $g^i_s \in [0.04, 0.12]$ (in $\text{S}/\text{m}$).
    \item The position $x_1\in\Omega$ and timing $\phi_0 \in [-25, 75]$ ms of the first activation site. In particular, we set $x_1$ from the UVCs $\theta_1 \in [-\pi, \pi]$, $\rho_1 \in [0, 1]$, and $z \in [0, 1]$.
\end{itemize}
Therefore, the dimension of the parameter space is 8.

The target ECG was obtained by taking the free parameters equal to mid-value in their range, and with the highest mesh resolution.

\subsubsection{Loss metrics}
\label{sec:employed_losses}


Besides standard error metrics, here we also consider dynamic time warping (DTW), see~\cite{sakoe_dynamic_1978}, as recently proposed by~\cite{camps_inference_2021}. In DTW, a cost matrix $C$ is generated using a distance measure between all possible samples of two timeseries. The DTW algorithm will then find an optimal warp path through the cost matrix that optimally aligns the two signals, minimizing the cost of the warp path.

In summary, we compared the following metric distances~$\D(\mathbf{V}_j,\hat{\mathbf{V}_j})$:
\begin{enumerate}
    \item Least-squares error: $\D_2(\mathbf{f},\mathbf{g}) = \sum_n (f_n-g_n)^2$.
    \item Dynamic Time Warping (DTW) 
    \begin{itemize}
        \item $\D_{\text{DTW},1}(\mathbf{f},\mathbf{g})$ with $C_{n,m} = \abs{f_n-g_n}$.
        \item $\D_{\text{DTW},2}(\mathbf{f},\mathbf{g})$ with $C_{n,m} = (f_n - g_m)^2$.
    \end{itemize}
    \item Cosine distance: $\D_{\text{cos}}(\mathbf{f},\mathbf{g}) = 1 - \frac{\left<\mathbf{f}, \mathbf{g} \right>}{\norm{\mathbf{f}} \norm{\mathbf{g}}}$.
\end{enumerate}
Note that in all cases, the losses of all leads are simply summed up and in the case of DTW, the three cost matrices of the three leads $C_{X/Y/Z}$ are similarly added and DTW is performed on the resulting cost matrix $C = C_X + C_Y + C_Z$.

\subsubsection{Visualization of the loss}
To showcase the complexity of the model w.r.t.~the parameter space, it is often beneficial to visualize the loss landscape.
For this, we follow~\cite{yang_application_2017}, where the authors
consider two randomly sampled (possibly orthogonal) vectors from the parameter space and sample across a predefined range.
This can be thought of as conceptually slicing the $d$-dimensional hypercube with a hyperplane and computing the losses on this hyperplane.
The main advantage of such an approach is that no dimensionality reduction is required, while the interdependence of parameters in the parameter space is not lost.

The mentioned landscape can then be computed by sampling the loss functional $\D: \R^d \to \R$ as 
\begin{equation}
    \mathcal{L}(a, b) = \D( \mathbf{x} + a \eta + b \delta),
    \label{eq:loss_2D}
\end{equation}
where $\mathbf{x}$ is the parameter vector of the considered ground-truth solution and $\D$ is the chosen distance measure.

\section{Results}
\label{sec:results}

To give an overview of the loss landscape, we sampled the hyperplane with $50 \times 50$ samples, at 7 different resolutions for 5 different random direction pairs (hyperplanes), resulting in a total of $87,500$ computed ECGs.

\begin{figure*}[htb]
    \centering
    \includegraphics[width=\linewidth]{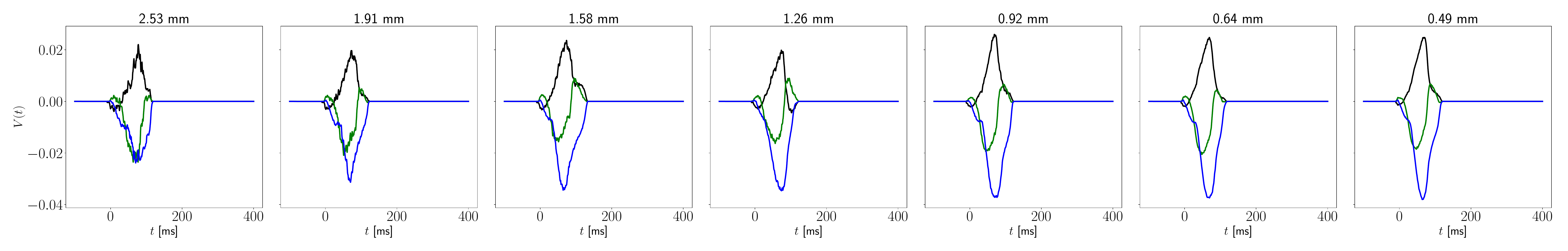}
    \caption{Plot of the computed ECGs for the optimal parameter configuration ($a = b = 0$ in~\eqref{eq:loss_2D}) at the tested different resolutions. 
    The ECGs get successively smoother, as we reduce the element size, but at a very slow rate}
    \label{fig:ecgs}
\end{figure*}

\begin{figure*}[htb]
    \centering
    \includegraphics[width=\linewidth]{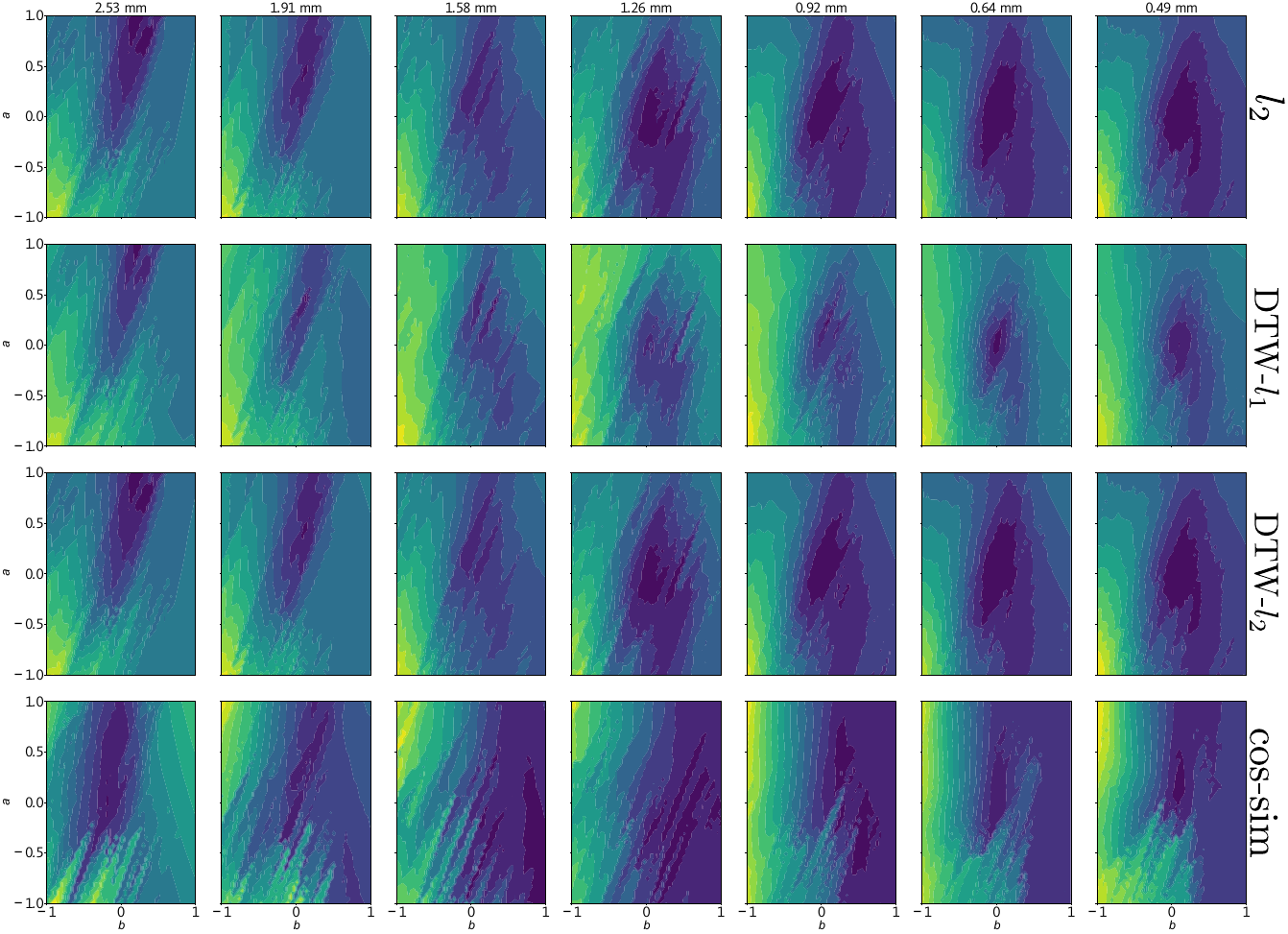}
    \caption{Loss landscapes over the whole tested parameter range at different resolutions and for multiple loss functions. 
    Colormap ranges are individual for each contour plot.}
    \label{fig:loss_landscapes}
\end{figure*}

We based our analysis on one of the random hyperplanes, but the overall conclusion drawn from all 3 hyperplanes remains the same.
We start by showing the computed ECG of the optimal configuration in Fig.~\ref{fig:ecgs} at different resolutions.
Note that a spatial resolution of $\approx 1$ mm is considered sufficient for most electrophysiological applications with the eikonal model, see~\cite{franzone_spreading_1993}. However, we still have some numerical errors on the signal.
This is most likely associated to the inaccuracy of capturing the very narrow wavefront of the integral~\eqref{eq:ecg}.
Such noise may be considered minor from a visual perspective, but is very likely responsible for the discontinuous loss landscapes as seen in Fig.~\ref{fig:loss_landscapes}.

The loss landscapes themselves also exhibit discretization artefacts especially on lower resolutions, most likely associated with the discretization of the variable initiation site, which is required to coincide with a vertex in our setup.
Perhaps surprisingly, nearly all loss functions perform similarly in terms of smoothness for the tested parameter ranges and no significant advantage can be seen by applying the more sophisticated DTW instead of the straight-forward $l_2$ error.
However, we note that in all parameter scenarios, the main peaks of the signals at least partially overlap.
When trying to match non-overlapping or only slightly overlapping signals, the advantage of DTW may be more pronounced.

It is noteworthy that we also calculated the loss after applying mean filters to the ECG that visually removed most of the noise, but did only change the smoothness of the loss landscape by an insignificant amount.

%

\section{Discussion}
\label{sec:disc}
This paper offers some insights on the impact and significance of parameters on different loss functions w.r.t.~the ECG.

The complete pipeline in Sec.~\ref{sec:methods} involves a several steps, some of which involve solutions of PDEs. 
From the viewpoint of optimization, it is important to understand the smoothness of the loss function w.r.t.\ its argument, that is in our case the parameters of the forward model.  
When the loss function itself is non-smooth, obviously we cannot expect parametric smoothness. This is the case of the $l^1$-error or the DTW.  
For the other error metrics, the smoothness of the loss function will in fact rely on the parametric smoothness of the ECG $V_j(t)$. 
In our experiments, the most restrictive parameters in terms of regularity are the conduction velocity tensor and the initiation site. 
Together, they define the eikonal model for computing the activation map.  
It is possible to show that the activation map is Lipschitz-continuous w.r.t.\ the initiation site (either location or onset timing).
Therefore, the gradient of the activation, parallel to the propagation direction and used in the computation of the ECG, is defined almost everywhere. 
As a matter of fact, in virtue of~\eqref{eq:ecg}, the singularity set has little effect on the ECG. A similar argument applies to the conduction velocity tensor, which depends on the fiber angles and the conductivity values.

More prominent is the role of the numerical discretization. 
In particular, the location of the initiation sites are Dirichlet boundary conditions to the eikonal equation~\eqref{eq:eikonal}, therefore applied at discrete locations (the mesh vertices). In such a case, a projection to the closest vertex in the mesh needs to be considered, resulting in a discontinuous loss function.

Another problem is that the temporal evaluation of the ECG should follow, to some extent, the spatial grid size.  
First, please note that in~\eqref{eq:waveform} the time is a parameter. 
For some fixed $t$, $v_m$ is generally interpolated on the computational grid (or, equivalently, evaluated at some quadrature nodes) for the computation of the ECG in~\eqref{eq:ecg}. 
Therefore, when the time step is too small, the narrow function $\nabla v_m$ might be poorly approximated when barely touching any degrees-of-freedom.
This translates into (numerical) oscillations on the surface ECG, as reported in Fig.~\ref{fig:ecgs}. 
The threshold value for the appearance of oscillations is related to the conduction velocity, because when the wave moves fast the \emph{saltatory} propagation disappears. 
In fact, the ratio between the mesh size and the time step should be comparable to the conduction velocity. 
This is unfortunately difficult sometimes, because the time step is a global parameter whereas mesh size and conduction velocity is local to the mesh.

The loss function itself does not appear to be very smooth in the parameter space, as deduced from the level sets (see Fig.~\ref{fig:loss_landscapes}). 
Possible sources of this noise are time discretization and spatial discretization (element size), where we suspect that~\eqref{eq:ecg} would especially benefit from higher order smoothness of the solution, or higher resolutions.
Additionally, the Lipschitz-continuity of $\phi$ translates to $v_m$, which could be remedied or at least mitigated by an eikonal-diffusion formulation, which has also been shown to be a closer approximation to the underlying physiological processes~\cite{keener_mathematical_1998}.

Some simplifying assumptions have been made in the course of this study to make the many computations more tractable:
For one, the lead field is usually not axis aligned, but rather a result of another elliptic Poisson PDE.
As the leads are usually far away from the computational domain (the heart), the impact however may not be very large.
Secondly, the eikonal solution is an approximation of the bidomain model, but is preferrable due to its low computational demand, while still maintaining model fidelity.

Still, the present study provides an initial step in better understanding the complexity of fitting models to given ECGs.
Such insights will be useful in future optimization and machine learning methods that rely and benefit from a smooth loss function to solve the electrophysiological inverse problem to ultimately bring us closer to cardiac digital twinning using a wide range of parameters.

\begin{ack}
We would like to thank Matthias Gsell for sharing insights on practical considerations for biventricular models, especially UVCs. 
\end{ack}

\bibliography{smooth}             
\listoftodos
\end{document}